\documentclass[useAMS,usenatbib]{mn2e}

\usepackage{graphicx}

\title[WASP-1: A lithium- and metal-rich star]
      {WASP-1: A lithium- and metal-rich star with
      an oversized planet\thanks{Based on observations made
      with the Nordic Optical Telescope.}}

\author[H. C. Stempels, A. Collier Cameron, L. Hebb, B. Smalley and S. Frandsen]
       {H. C. Stempels$^{1}$\thanks{E-mail: Eric.Stempels@st-andrews.ac.uk}
        A. Collier Cameron$^{1}$,
	L. Hebb$^{1}$,
	B. Smalley$^{2}$,
	and S. Frandsen$^{3}$\\
$^{1}$School of Physics \& Astronomy, University of St Andrews, North Haugh, St
Andrews, Fife, KY16 9SS, UK\\
$^{2}$Astrophysics Group, Keele University, Staffordshire, ST5 5BG, UK\\
$^{3}$Dept. of Physics and Astronomy, {\AA}rhus University, Denmark
}
\begin{document}

\date{Accepted XXXX. Received XXXX; in original form XXXX}

\pagerange{\pageref{firstpage}--\pageref{lastpage}} \pubyear{2007}

\maketitle

\label{firstpage}

\begin{abstract}
In this paper we present our results of a comprehensive spectroscopic analysis
of WASP-1, the host star to the exoplanet WASP-1b. We derive $T_{\rm eff} =
6110 \pm 45$ K, $\log g = 4.28 \pm 0.15$, and [M/H] $=0.23 \pm 0.08$, and also a
high abundance of lithium, $\log n({\rm Li}) = 2.91 \pm 0.05$. These parameters
suggests an age for the system of 1--3~Gyr and a stellar mass of
1.25--1.35 $M_{\odot}$.
This means that WASP-1 has properties very similar to those of HD~149026,
the host star for the highest density planet yet detected. Moreover, their
planets orbit at comparable distances and receive  comparable irradiating fluxes
from their host stars. However, despite the similarity of WASP-1 with
HD~149026, their planets have strongly different densities. This suggests that
gas-giant planet density is not a simple function of host-star metallicity or of
radiation environment at ages of $\sim 2$ Gyr.
\end{abstract}


\section{Introduction}

Close-orbiting giant exoplanets that transit their parent stars offer unique
insights into the interior structure and evolution of gas-giant planets. They
are the only extrasolar planets whose radii and masses can be determined
unambiguously from their light curves and radial-velocity variations. The
discovery of the first such planet, HD 209458b \citep{charbonneau00,henry00},
established its gas-giant nature. Since then, 13 other transiting planets have
been discovered, revealing a diverse range of densities and hence internal
compositions. The Saturn-mass planet HD 149026b represents one extreme, having a
high density requiring a substantial rock/ice core. At the other end of the
density scale one finds HD 209458b, HAT-P-1b \citep{bakos07} and WASP-1b
\citep{CC07}. These planets have densities so low that even coreless models have
difficulty reproducing their radii \citep{fortney06,burrows06}.
\citet{guillot06} pointed out that the core masses inferred for then-known
transiting planets show a weak correlation with the metallicity of the parent
star. \citet{burrows06} find a similar trend. They also noted that
\mbox{WASP-1b}, \mbox{WASP-2b} and \mbox{XO-1b} \citep{mccullough06} have very
similar masses but substantially different radii. In the absence of detailed
spectroscopic abundance analyses for WASP-1 and WASP-2, however, it is
difficult to determine a likely cause for 
their inflated radii.

WASP-1b has, in
comparison with other planets, a very low density; it is 2.4 times as massive as
HD 149026b, but also 3.3 times less dense. This poses a significant challenge to
existing theoretical models. \citet{burrows06} find it hard to
reconcile the planet's large radius $R_p=1.44 R_{\rm Jup}$
\citep{charbonneau07,shporer07} with evolutionary tracks even for coreless
models with ten-times-solar atmospheric abundances and ages greater than
1.5 Gyr. This leads to a wide range of speculations. Could it be possible
that the star (and by implication its planet) is relatively young, and thus
still contracting? Could the strong irradiation of the planet inhibit
contraction \citep{guillot02}? Can additional interior opacity due to enhanced
atmospheric metallicity slow contraction \citep{burrows06}? Or perhaps (as a
counter-argument) can an increased interior molecular weight due to enhanced
atmospheric metallicity accelerate contraction? Can tidal heating caused by
orbital eccentricity \citep{bodenheimer01} or rotational obliquity
\citep{winn05} provide additional interior pressure support? Given so
many theoretical candidate mechanisms for inflating a planet's radius,
comparative studies of the rapidly-growing number of stars that host
well-characterised transiting planets are essential if we are to identify the
dominant environmental and evolutionary processes that determine a  mature
gas-giant planet's internal structure and outer radius.

Here we present a detailed analysis of the spectrum of the parent star to
WASP-1b, WASP-1 (= GSC 02265--00107), with the goal of establishing its
metallicity and evolutionary status. We use as our starting point the
preliminary stellar parameters published by \citet{CC07}. Although their
analysis is roughly similar to the one presented here, the spectra they obtained
from the SOPHIE spectrograph at OHP were compromised by background
scattered-light contamination and uncertain continuum normalization, leaving the
metallicity unconstrained. We also compare the metallicity, effective
temperature and evolutionary age to those of the very similar host star of
HD~149028b, and discuss whether the properties of the host stars give any useful
clues to the strongly-contrasting densities of their close-orbiting gas-giant
planets.

\section{Observations and data reduction}

\begin{figure}
  \centering
  \includegraphics[angle=90,width=\columnwidth]{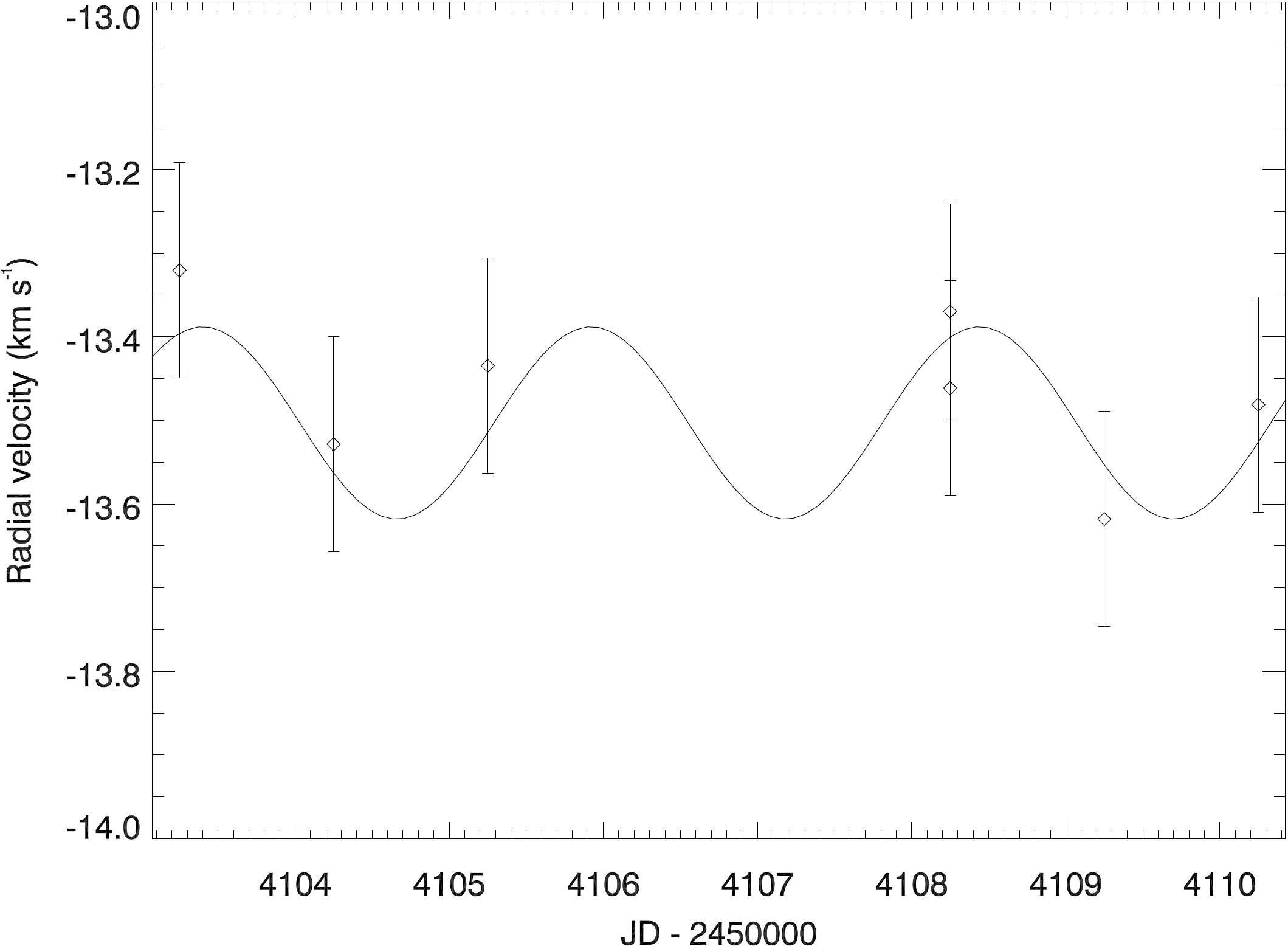}
  \caption{Radial velocity measurements from the seven obtained spectra of
  WASP-1, compared with the ephemeris determined by \citet{CC07}.}
  \label{fig:radvel}
\end{figure}

WASP-1 was observed in early January 2007 at the 2.5m Nordic Optical Telescope
(NOT), as part of the science verification program of the newly commissioned
spectrograph FIES (FIber Echelle Spectrograph). We obtained a total of seven
spectra on six nights, with a spectral resolution of $R=47\,000$ and an exposure
time of 20 minutes each, covering the wavelength region $4000$--$7350$\,{\AA} at
a signal-to-noise of 40--50 in each spectrum.

The FIES spectrograph is a bench-mounted fiber-fed echelle spectrograph, located
in its own temperature-stabilized building and without any moving parts. With
such a particular setup one obtains very stable spectra in a fixed format for
which one of the authors (HCS) developed an automated data reduction system.
This system uses {\sc Python} and {\sc PyRAF} to access the echelle reduction
routines of {\sc IRAF}. The reduction is fine-tuned for the properties of FIES,
and performs all necessary reduction steps, such as subtraction of biases and
scattered light, flat-fielding, order extraction, normalization (including
fringe-correction) and wavelength calibration, leaving the observer with
fully-reduced spectra ready for scientific analysis.

Using the associated wavelength calibration frames we corrected each spectrum
for instrumental shifts. We also applied the correction for the heliocentric
velocity and checked and calibrated the spectra against a radial velocity
standard. For the purpose of determining the stellar parameters of WASP-1, we
also created one high-quality spectrum with excellent signal-to-noise ($\sim
100$) by combining the seven individual spectra into one.

\section{Properties of WASP-1}

\subsection{Radial velocities}

The seven individual spectra we obtained provide an independent check of the
sinusoidal radial-velocity variations detected in WASP-1 by \citet{CC07}. In
order to obtain the best accuracy, we cross-correlated high signal-to-noise
regions of the spectra on an order-by-order basis. In this analysis we excluded
orders that contain atmospheric features. This allowed us to determine radial
velocities with an accuracy of $\pm 125$ m s${}^{-1}$. We find that the
ephemeris determined by \citet{CC07} is compatible with our observations (see
Fig. \ref{fig:radvel}).

\begin{table}
  \begin{minipage}{\columnwidth}
  \centering
  \caption{Obtained parameters for WASP-1. See
  Sect.~\ref{sec:errors} for a description of how the uncertainties were
  derived.}
  \begin{tabular}{lr@{ $\pm$ }l r@{ $\pm$ }l}
  \hline
  Parameter	& \multicolumn{2}{c}{Value} & \multicolumn{2}{c}{Correlated} \\
  & \multicolumn{2}{c}{} & \multicolumn{2}{c}{uncertainty} \\
  \hline
  $T_{\rm eff}$	& $6110$ & $45$	K	& & 50 K	 \\
  $\log g$ 	& $4.28$ & $0.15$	& & 0.05	 \\
  {[M/H]}	& $0.23$ & $0.08$	& & 0.03	 \\
  $\log n({\rm Li})$ & $2.91$ & $0.05$	& & 0.05	 \\
  {[Na/H]}  	& $ 0.12$ & $0.08$	& & 0.02	 \\
  {[Si/H]}  	& $ 0.26$ & $0.06$	& & 0.01	 \\
  {[Ti/H]}  	& $ 0.30$ & $0.16$	& & 0.01	 \\
  {[Fe/H]}  	& $ 0.26$ & $0.03$	& & 0.02	 \\
  {[Ni/H]}  	& $ 0.24$ & $0.07$	& & 0.02	 \\
  {[Mg/H]}  	& $ 0.10$ & $0.05$	& & 0.04	 \\
  $v \sin i$	& $ 5.79^*$ & $0.35$ km\,s${}^{-1}$ & \multicolumn{2}{c}{--} \\
  $v_{\rm rad}$	& $-13.46$ & $0.1$ km\,s${}^{-1}$ & \multicolumn{2}{c}{--} \\
  \hline
  \multicolumn{5}{l}{$^*$ Assuming a fixed value for the macroturbulence.}\\
  \multicolumn{5}{l}{Otherwise an upper limit (see Sect. \ref{sec:errors})}\\
  \end{tabular}
  \end{minipage}
  \label{tab:params}
\end{table}

\begin{figure*}
  \centering
  \includegraphics[angle=90,width=15cm]{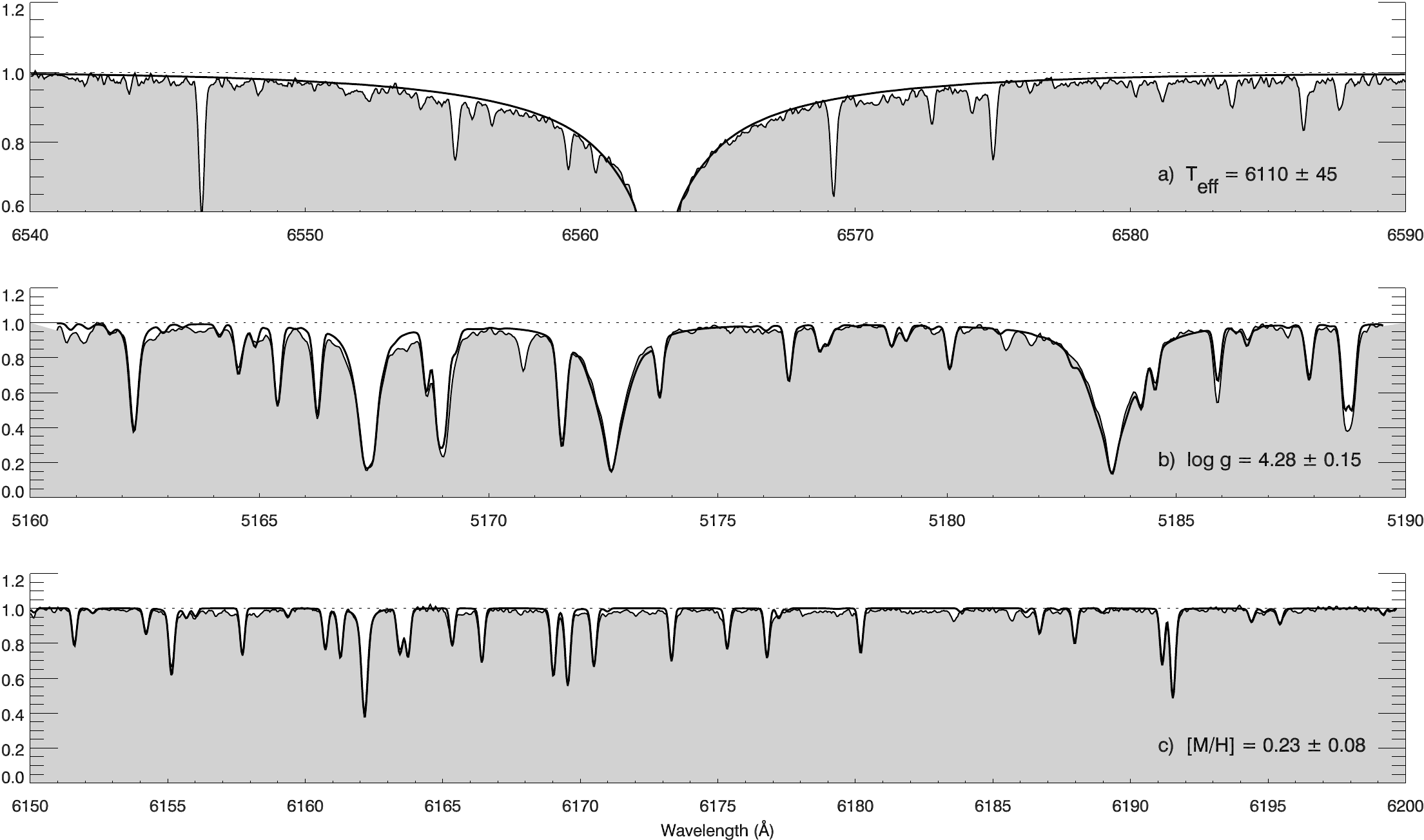}
  \caption{The above panels show a comparison of the observed spectrum (grey)
  with the synthetic spectrum based on the recovered parameters (solid, thick
  line). Panel a) shows the temperature-sensitive H$\alpha$ 6563 line, b) the
  Mg\,b 5175 triplet, sensitive to $\log g$ c) a section of the region
  containing a wealth of metal lines, sensitive to [M/H] and d) the Li {\sc i}
  6708 line. Thes stellar parameters of WASP-1 were not determined from each
  panel individually, but from all spectral regions simultaneously (see text).
  }
  \label{fig:profs}
\end{figure*}

\subsection{Spectroscopic analysis}
\label{sec:specanal}

Several detailed spectroscopic investigations of exoplanet hosts have been
presented in the literature \citep[among
others,][]{gonzalez01,VF05,santos04,santos06}. The primary difference between
these studies is the choice of numerical methods and model atmospheres to
determine which stellar parameters best describe the stellar spectrum. In this
paper we choose to follow the methodology of one of these studies, namely
\citet[hereafter VF05]{VF05}, which was also used in the studies of the
transiting exoplanet systems XO-1 \citep{mccullough06} and HAT-P-1
\citep{bakos07}. We make identical assumptions and use of the same tools,
techniques and grid of model atmosheres, allowing us to obtain stellar
parameters for \mbox{WASP-1} that can be directly compared to the results of
VF05.

We analysed our high-quality spectrum of WASP-1 with {\sc SME} \citep[{\sc
Spectroscopy Made Easy}, see][]{valenti96}, an {\sc IDL}-based program that 
uses synthetic spectra and multi-dimensional least-squares minimization to
determine the best set of stellar parameters (the effective temperature $T_{\rm
eff}$, the gravity $\log g$, the metallicity [M/H], the projected radial
velocity $v \sin i$, the systemic radial velocity $v_{\rm rad}$, the
microturbulence $v_{\rm mic}$ and the macroturbulence $v_{\rm mac}$) that
describe an observed spectrum.

Successful calculation of synthetic spectra requires a grid of model atmospheres
covering the parameter space of interest, as well as accurate line lists of
atomic transitions. In analogy with VF05, we used for our analysis the routine
for 3-dimensional interpolation on the \citet{kurucz93} grid of LTE model
atmospheres. Atomic line data was obtained from the VALD database
\citep{piskunov95,kupka99}. To improve our ability to model the stellar spectrum
in the spectral regions of interest (see below), we adjusted the oscillator
strengths and broadening parameters for some of the lines in our line lists.
This was done by comparing the NSO spectrum of the Sun \citep{kurucz84} to a
synthetic spectrum (using the parameters $T_{\rm eff} = 5770$ K, $\log g =
4.44$, $v \sin i = 1.4$ km\,s${}^{-1}$, $v_{\rm rad} = 0.4$ km\,s${}^{-1}$
(gravitational blueshift), $v_{\rm mic} = 0.866$ km\,s${}^{-1}$ and $v_{\rm mac}
= 3.57$ km\,s${}^{-1}$ and solar abundances). We also checked our results
against a high-quality spectrum of Procyon, obtained with FIES at the NOT
\citep[for stellar parameters, see][]{fuhrmann97}.

In order to use assumptions identical to those of VF05, we decoupled the
correlation between microturbulence $v_{\rm mic}$ and metallicity by fixing the
value of $v_{\rm mic}$ to 0.85 km\,s${}^{-1}$. Similarly we followed their
empirical relation for the value of the macroturbulence, giving $v_{\rm mac} =
4.5$ km\,s${}^{-1}$ for a star with $T_{\rm eff} \approx 6200$ K (see also
Sect. \ref{sec:errors} for additional remarks on the choice of
macroturbulence). We also released the elemental abundances of five elements
(Na, Si, Ti, Fe and Ni). As an initial guess we used the parameters determined
by \citet{CC07}, as well as [M/H] $ = 0$, $v \sin i = 0.5$ km\,s${}^{-1}$, and
(Fe/H) $= -4.50$.

To constrain the full set of stellar parameters we identified three important
wavelength regions ($5160$--$5190$\,{\AA}, $6000$--$6200$\,{\AA} and
$6540$--$6590$\,{\AA}). The first region contains the triplet of Mg\,b lines,
primarily sensitive to $\log g$, while the second region contains a large number
of well-isolated and unresolved spectral lines of a range of different elements,
sensitive to $v_{\rm rad}$, $v \sin i$, [M/H], and the individual elemental
abundances in particular.

The third region covers the broad H$\alpha$ 6563 line. Although this region was
not included by VF05, we consider this region as a strong indicator to constrain
effective temperature. The broadening of the outer wings of this line is
sensitive to $T_{\rm eff}$, and largely independent of other stellar parameters.
Our synthetic synthesis calculations use the state-of-the-art hydrogen line
broadening theory of \citet{barklem00}. \citet{barklem02} showed that this
theory can accurately reproduce the Balmer line wings of stars with a range of
different spectral types, and that it produces good agreement with the results
of the infrared flux method (IRFM) developed by \citet{blackwell80}, but with
smaller error margins\footnote{While LTE simulations of Balmer line wings are
widely used to determine stellar temperatures, \citet{barklem07} recently
pointed out that it is unclear whether the assumption of LTE is applicable to
the Balmer lines, and that non-LTE calculations might yield temperatures
up to 100 K higher.}. They also showed that there is a weak dependence on the
model atmosphere used. 

During our analysis we encountered difficulties in obtaining a good fit to the
wings of the Mg\,b lines, sensitive to changes in $\log g$. While we obtained
good agreement with other $\log g$-sensitive spectral features (notably the
wings of Ca {\sc i} 6122 and Ca {\sc i} 6162), as well as with the spectra of
the Sun and Procyon, the synthetic profiles for the Mg\,b lines remained
too wide. Determining $\log g$ from the Mg b lines alone yielded a very
low value of $\log g = 3.95$, which is strongly inconsistent with the
photometrically determined radius of WASP-1. However, this apparent
inconsistency can be reconciled by adjusting the elemental abundance of [Mg/H].

Because our selected wavelength regions do not contain Mg lines other than
Mg b, we determined the Mg abundance in WASP-1 from seven weak Mg lines in other
regions of the spectrum (Mg~{\sc i}~5528.4, 5711.1, 5785.3, 5785.6, 6318.7, 
6319.2 and 6319.5 {\AA}). We find [Mg/H]$ = 0.10$, indicating that Mg is indeed
underabundant by $0.12$ in \mbox{WASP-1}. Similar underabundances in Mg (and Na)
for metal rich stars and exoplanets hosts was reported by \citet{gonzalez01},
although other comprehensive studies by \citet{bensby05} and \citet{gilli06} do
not confirm such a trend. We then redetermined the stellar parameters of
WASP-1, treating [Mg/H] as a fixed parameter. This produced only minor changes
in the derived stellar parameters, but also a good agreement between the
observed and synthetic profiles around the Mg b line complex.

The final set of parameters we obtained from the analysis with {\sc SME} are
listed in Table~\ref{tab:params}. We also reproduce the synthetic spectra of the
best choice of parameters in Fig~\ref{fig:profs}. We would like to point out
that all parameters were determined simultaneously across all three wavelength
regions, thus obtaining a self-consistent solution.

\subsection{Error estimates and parameter correlations}
\label{sec:errors}

As VF05 point out, the formal uncertainties reported by {\sc SME} strongly
underestimate the true uncertainties because the value of the goodness-of-fit
parameter $\chi^2$ is not dominated by the signal-to-noise of our spectra, but
rather by uncertainties in the data reduction (for example, continuum fitting)
as well as by inaccuracies in the synthetic spectra calculations. Thus, in
analogy with VF05, we individually analysed our seven spectra of \mbox{WASP-1}
with SME, and quote the square root of the variance of the measurements of each
parameter as the uncertainty in Table~\ref{tab:params}.

Spectroscopically determined fundamental parameters tend to show strong
correlations, and any uncertainty in one parameter may be propagated to other
parameters. We investigated the magnitude of these correlations by varying
$T_{\rm eff}$ with $\pm 50$~K, and determining a new optimal solution at these
temperatures, thus propagating the uncertainty in temperature to the other
parameters.
The correlated uncertainties (listed in Table \ref{tab:params}) indicate
that the uncertainties due to parameter correlations are of similar
order of magnitude as the uncertainties derived earlier from the scatter of
measurements of the seven individual spectra.

In addition to propagating the uncertainties related to temperature, we also
looked at the relation between [Mg/H], [M/H] and $\log g$, as these may be
expected to correlate through the Mg b line complex. We find a clear correlation
between [Mg/H] and [M/H], such that [Mg/H] always shows a relative
underabundance of about $-0.1$ with respect to [M/H]. On the other hand we see
no indication that the value of $\log g$ correlates with [Mg/H], mainly because
each parameter is well constrained by diagnostics other than Mg b (i.e. the weak
Mg lines and the broadening of Ca {\sc i} 6122 and Ca {\sc i} 6162).

One pair of parameters that shows a strong correlation are the assumed value of
the macroturbulence $v_{\rm mac}$ and the obtained projected rotational velocity
$v \sin i$. This is no surprise, because the observed photospheric line profiles
are a convolution of rotational and turbulent broadening. In our analysis we
have assumed a value of 4.5 km\,s${}^{-1}$, similar to the choice of VF05. This
value is in agreement with several other empirical relations for main-sequence
type stars (see VF05). However, the parameters we derive for WASP-1 could
indicate that WASP-1 is starting to evolve off the main-sequence (see Sect.
\ref{sec:evol}), which would imply a slight increase in $v_{\rm mac}$, to
approximately $ 5.5$ km\,s${}^{-1}$. Using higher values for $v_{\rm mac}$ would
result in a lower value of the $v \sin i$, thus the value we quote in Table
\ref{tab:params} should be considered an upper limit.

\subsection{Independent estimate of $T_{\rm eff}$ from the IRFM}

In addition to the spectroscopic estimate, we also determined the effective
temperature using the IRFM. This method allows for a
nearly model-independent determination of $T_{\rm eff}$ obtained from the 
integrated stellar flux at the Earth and a measurement at an infrared wavelength.

For WASP-1 we used the Tycho $B$ and $V$ magnitudes and the 2MASS magnitudes to
estimate the integrated flux ($F_\oplus$) by fitting Kurucz flux
distributions. Since WASP-1 is relatively nearby and out of Galactic plane we
assumed zero reddening. Integrating the fitted distributions gives a value of 
$F_{\rm tot} = (6.2 \pm 0.5) \cdot 10^{-13}$ W m${}^{-2}$. Using the IRFM to determine the
$T_{\rm eff}$ for the $J$, $H$ and $K_s$ bands, we obtain an average $T_{\rm eff} =
6200 \pm 200$ K, which is in close agreement with the spectroscopic
determination.

\begin{figure}
  \centering
  \includegraphics[angle=90,width=\columnwidth]{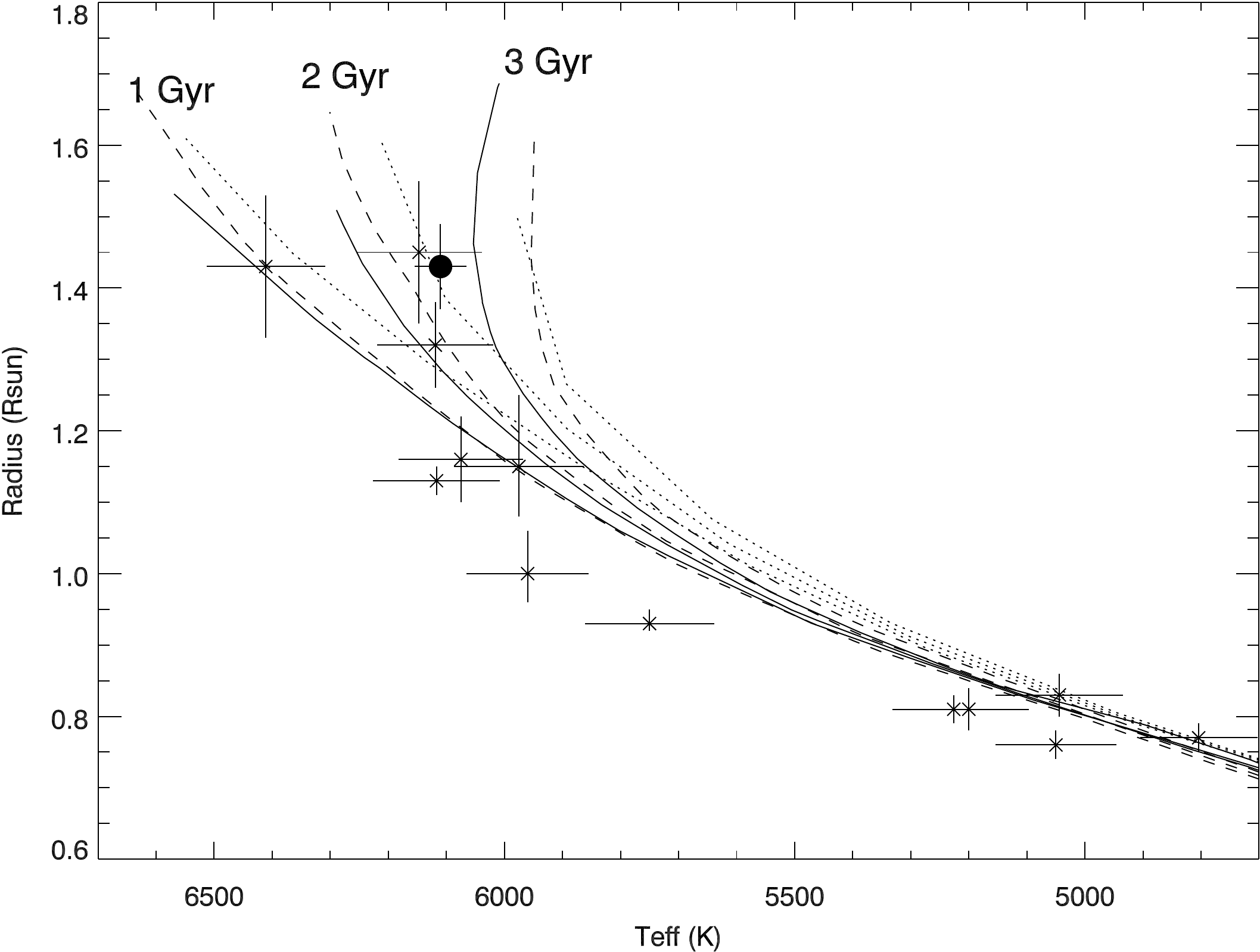}
  \caption{Radius versus $T_{\rm eff}$ (a modified Hertzprung-Russel diagram)
  of the host stars of the 14 known transiting extra-solar planets
  \citet[asterisks, ][]{burrows06}, including \mbox{WASP-1} (solid circle). 
  Here we present isochrones for super-solar metallicity stellar evolution
  models from \citet{girardi02} (solid), \citet{yi02} (dotted), and
  \citet{siess97} (dashed). Given the uncertainties inherent in the models from
  various authors, the position of WASP-1 is consistent with an age of
  \mbox{1--3~Gyr}.}
  \label{fig:hrd}
\end{figure}

\section{Evolutionary status}
\label{sec:evol}

The stellar parameters derived from the detailed spectroscopic analysis
described above can be used to place \mbox{WASP-1} in its evolutionary context, 
providing further constraints on the age of the host star and its planet.
Both the effective temperature and the (photometrically determined)
stellar radius are measured to high accuracy using well understood, robust
physics that is largely model independent. Thus in Fig.~\ref{fig:hrd}, we plot a
modified Hertzprung-Russel diagram in which the radius acts as a proxy for
luminosity (as it is related to luminosity through the temperature) and compare
the properties of WASP-1 to several different super-solar metallicity
stellar evolution models \citep{girardi02,siess97,yi02} between 1--3~Gyr. The
position of WASP-1 in this diagram shows that its its age is likely to be in the
range 1--3 Gyr and that its large stellar radius may be explained by the fact
that the star is beginning to evolve off the main sequence.

We note that the mass initially adopted for WASP-1, $M=1.15~M_{\odot}$,
\citep{CC07} is marginally inconsistent with the $\log g$ found here, when
combined with the
latest estimates of the stellar radius
\citep[$R_* = 1.453 \pm 0.032 R_{\cdot}$][]{charbonneau07,shporer07}. A
slightly higher mass of $M=1.25$--$1.35~M_{\odot}$ would allow all the measured
values of $\log g$, temperature, and radius to be consistent given their errors.

\subsection{Lithium abundance}

An interesting feature in the spectrum of WASP-1 is the the strong lithium
absorption line at $6708$\,{\AA}. As stars only destroy lithium, the amount of
photospheric lithium is often used as an age indicator. However, the accuracy of
age estimates based on lithium abundances depends strongly on the efficiency
with which stars deplete lithium.
Stars such as WASP-1, with temperatures of $\sim 6100$ K, cannot
deplete lithium very easily from their surface layers, because the bottom of
the convection zone does not reach temperatures high enough to burn lithium. It
is also cooler than 6300 K, the lower limit of the so-called ``Li dip'' or
``Boesgaard gap'' \citep[for example,][]{balachandran95} where lithium depletion
is enhanced by an as yet unknown process. Still, early type stars do
slowly deplete lithium during their main sequence lifetimes \citep{jones99},
allowing modest constraints to be placed on the age on WASP-1 based on the
abundance of lithium present in the photosphere.

\begin{figure}
  \centering
  \includegraphics[angle=90,width=\columnwidth]{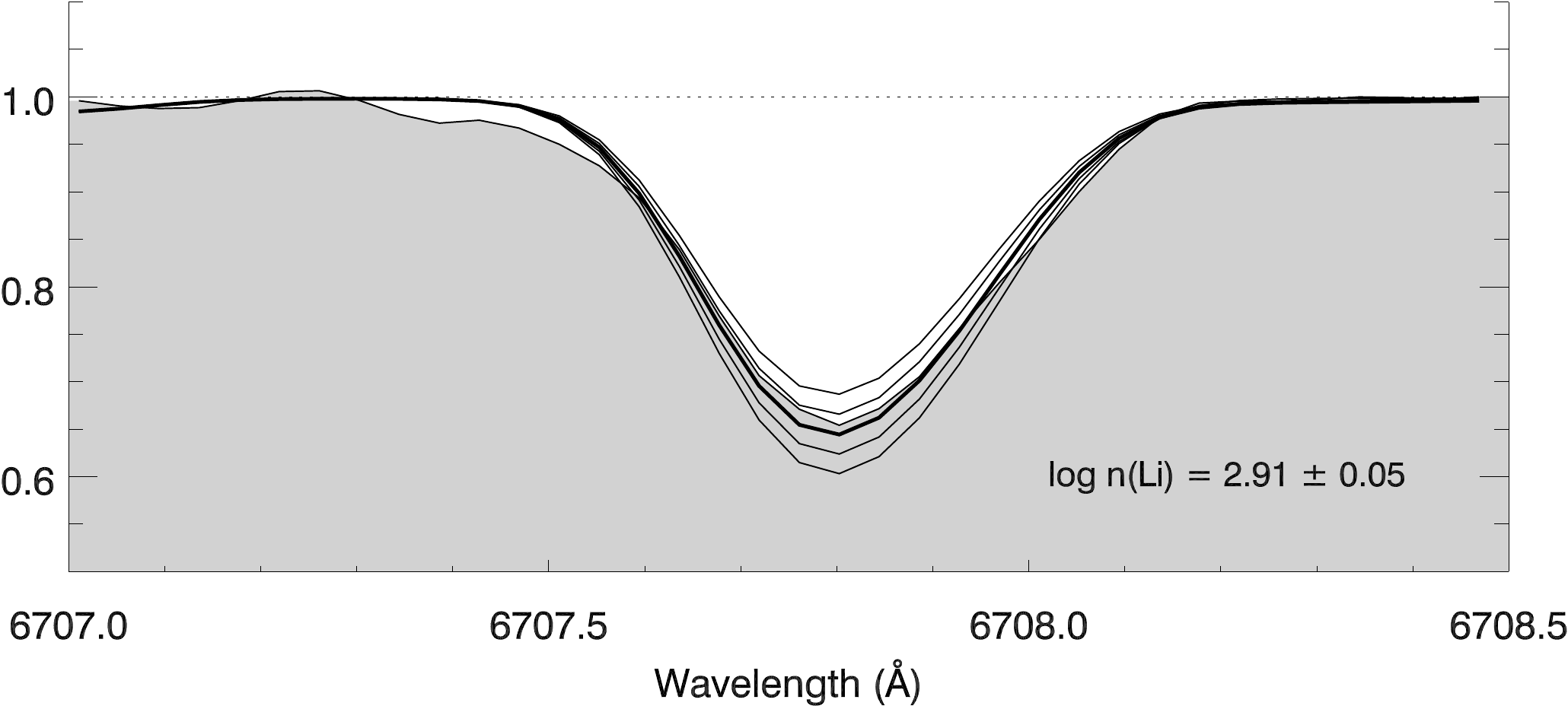}
  \caption{A range of synthetic line profiles for the Li {\sc i} 6708 feature,
  in steps of $0.05$ in $\log n({\rm Li})$. The thick line corresponds to the
  optimum value, $\log n({\rm Li}) = 2.91$. Changes of $\pm 50$ K in the
  stellar effective temperature result in changes of $\pm 0.05$ in the lithium
  abundance (see text).}
  \label{fig:liplot}
\end{figure}

In the spectrum of WASP-1, Li {\sc i} 6708\ has an equivalenth width of
$135$ m{\AA}. Using {\sc SME} and the atmospheric parameters derived
earlier, we find that this corresponds to an elemental abundance of ${\rm
(Li/H)} = -9.09 \pm 0.05$, or $\log n({\rm Li}) = {\rm (Li/H)} + 12 = 2.91 \pm
0.05$ in more conventional units (see Figure \ref{fig:liplot}). We find a
weak correlation between $T_{\rm eff}$ and $\log n({\rm Li})$, where a decrease
in temperature by 50 K relates to a decrease in $\log n({\rm Li})$ by 0.05.

The value $\log n({\rm Li}) = 2.91 \pm 0.05$ is less than the commonly
adopted value of 3.2 for the primordial lithium abundance. Still, it is a the
high end with respect to the study by \citet{israelian04}, who compared the
amount of lithium in exoplanet host stars with a set of reference stars.
We also compared our value of $\log n({\rm Li})$ with the empirical relations of
lithium depletion in open clusters derived by \citet{sestito05}. Their study
shows that, for stars with temperatures of $6200 \pm 150$ K, lithium can be
expected to reach $\log n({\rm Li}) = 2.9$ after 1--3 Gyr. This age estimate is
consistent with what we derived earlier from evolutionary tracks.

\section{Discussion}

Our spectroscopic analysis, together with the stellar mass and radius estimates
inferred from previous studies, reveal a remarkable similarity between the
physical parameters of WASP-1 and HD~149026 (see Table \ref{tab:compare}.
The two stars appear to be very similar in both mass and evolutionary status.
\citet{sato05} estimate the age of HD 149026 to be $2.0\pm 0.8$~Gyr.
Also the environments in which the two planets orbit are remarkably
similar. WASP-1b orbits slightly closer to a slightly cooler star, with the
result that its blackbody effective temperature (assuming a low albedo and
efficient redistribution and reradiation of heat from the night side) is just
80~K warmer than that of HD~149026b.

The structural differences between the two planets contrasts starkly with
the similarities of the parent stars. The light, Saturn-mass HD 149026b 
requires a core mass of several tens of Earth masses to reproduce its small
radius and high density \citep{sato05,fortney06}. WASP-1b is more than
twice as massive and has a much larger radius. This can only be
reconciled with the models of \citet{burrows06} if it has little or no core, its
atmosphere is metal-rich, and its age is less than 1.5 Gyr.

\begin{table}
  \begin{minipage}{\columnwidth}
  \centering
  \caption{A comparison between the stellar and planetary parameters of WASP-1
  and HD~149026. Data for HD~149026 are taken from \citet{charbonneau06} and
  \citet{sato05}. Data for WASP-1 are from \citet{CC07} and
  \citet{charbonneau07}, except $T_{{\rm eff},*}$, [Fe/H] and the age estimate,
  which are from this paper. The values of $T_{{\rm eff},p}$ were determined
  assuming  zero albedo and full redistribution of heat to the nightside.}
  \begin{tabular}{l l r@{}l r@{}l}
  \hline
  & & \multicolumn{2}{c}{HD149026} & \multicolumn{2}{c}{WASP-1} \\
  \hline
  $M_*$			& ($M_{\odot}$)	& $1.3~$ & $\pm~0.1$		& $1.15\,$ & ${}^{+0.24}_{-0.09}$ \\
  $R_*$			& ($R_{\odot}$)	& $1.45~$ & $\pm~0.1$		& $1.453~$ & $ \pm~0.032$\\
  $T_{{\rm eff},*}$	& (K)		& $6147~$ & $\pm~50$ K		& $6110~$ & $\pm~45$ K\\
  {[Fe/H]} 		&		& $0.36~$ &$\pm~0.05$		& $0.26~$ & $\pm~0.03$\\
  Age			& (Gyr)		& $2.0~$ & $\pm~0.8$		& $2.0~$ & $\pm~1.0$\\
  $M_p$			& ($M_{\rm Jup}$) & $0.330~$ & $\pm~0.002$	& $0.79\,$ & ${}^{+0.13}_{-0.06}$\\
  $R_p$			& ($R_{\rm Jup}$) & $0.726~$ & $\pm~0.064$	& $1.443~$ & $\pm~0.039$\\
  $\rho_p$		& ($\rho_{\rm Jup}$) & \multicolumn{2}{c}{$0.86$}& \multicolumn{2}{c}{$0.26$}\\
  orbital sep.		& (AU)		& \multicolumn{2}{c}{$0.042$}	& $0.0379~$ & $\pm~0.0042$\\
  $T_{{\rm eff},p}$	& (K)		& \multicolumn{2}{c}{$1740$}	& \multicolumn{2}{c}{$1820$}\\
  \hline
  \end{tabular}
  \end{minipage}
  \label{tab:compare}
\end{table}

Although HD 149026 could be a factor 2 more metal-rich than WASP-1 within
the uncertainties, WASP-1 itself is substantially more metal-rich than the Sun.
With this caveat in mind, we suggest that these two planets provide a strong
counter-example to the trend suggested tentatively by \citet{guillot06} and
\citet{burrows06}, in which core mass increases with metallicity of the host
star. We conclude that the final balance between core and envelope mass in giant
planets cannot be a simple function of stellar metallicity, nor of
radiation environment. As \citet{ikoma06} conclude, the final composition of a
planet is more likely to be dictated by the details of the disk environment in
which the planet formed, and possibly the dynamical history following the
envelope accretion phase, than by the composition of the parent star.

\section*{Acknowledgments}
We would like to thank the referees for constructive critisism that helped to
improve this paper. HCS acknowledges support from the Swedish Research Council.
The Nordic Optical Telescope is operated on the island of La Palma jointly by
Denmark, Finland, Iceland, Norway, and Sweden, in the Spanish Observatorio del
Roque de los Muchachos of the Instituto de Astrofisica de Canarias.

\bsp

\label{lastpage}

\end{document}